\documentstyle[pre,aps,epsf]{revtex}
\input epsf
\newcommand{\be}{\begin{equation}}
\newcommand{\ee}{\end{equation}}
\newcommand{\bea}{\begin{eqnarray}}
\newcommand{\eea}{\end{eqnarray}}

\begin{document}
\draft
\title{Calculations of exchange interaction in impurity band of two-dimensional semiconductors with 
out of plane impurities}
\author{I. V. Ponomarev$^{a}$, V. V. Flambaum}
\address{School of Physics, The University of New South Wales,
Sydney 2052, Australia}
\author{A. L. Efros}
\address{Department of Physics, University of Utah, Salt Lake City UT 84112}
\date{\today}
\maketitle
\begin{abstract}
We calculate the singlet-triplet splitting for a couple of two-dimensional electrons in
the potential of two positively charged impurities which are located out of plane. We consider
different relations between vertical distances of impurities $h_1$ and
$h_2$ and their lateral distance $R$. Such a system has never been studied in atomic physics but
the methods, worked out for regular two-atomic molecules and helium atom, have been found to be
useful.  Analytical expressions for several different limiting configurations of impurities
are obtained an interpolated formula for intermediate range of parameters is proposed. The
$R$-dependence of the splitting is shown to become weaker with increasing $h_1,h_2$.
\end{abstract}

 
\section{Introduction}
Two-particle exchange interaction is the main mechanism of the spin-spin interaction 
in the impurity band of semiconductors. 
The recent experiments\cite{Krav1,Krav2} on the metallic phase in 
two-dimensional 
system show that a  magnetic field parallel to the plane of the two-dimensional electron gas
can  destroy this phase. This suggests that the
spin interaction plays an important role in the formation of the metallic state. 
 The study of low-dimensional spin systems is usually based on the Heisenberg
Hamiltonian
$\sum_{\langle i k \rangle}J_{ik}{\bf s_i\cdot s_k}$,
where ${\bf s}$ is the  spin 1/2 operator,  and $i,k$ denote 
different spins.  Thus, the   knowledge of the exchange constants $J_{ik}$, which 
are  equal to half of the singlet-triplet
energy splitting for two electrons located at different sites is very important.

In our previous paper\cite{pon} we have calculated these parameters and have obtained analytical
expression for the ground state and excited states of the Heisenberg Hamiltonian at small impurity
density. We assumed, however that impurities are located in the same  plane as electron gas. 
This
is not the case in usual semiconductor structures, where impurities are separated from electrons
by the so-called spacer layer and residual impurities are also out of plane containing
electron gas. In this paper we concentrate on the problem of spin splitting in the system of
two electrons bound to the Coulomb centers which are located outside the plain.

Fig. 1. illustrates the general geometry of the problem. 
\begin{figure}\label{fig1}
\epsfxsize=14 truecm
\epsfysize=7 truecm
\centerline{\epsffile{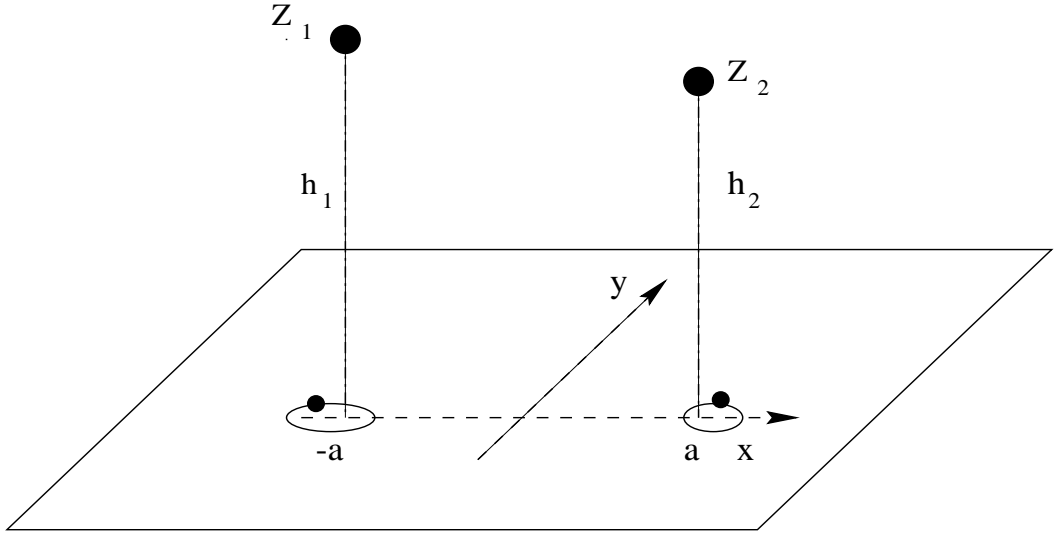}}
\caption{  General geometry of the problem }
\end{figure}
 The total Hamiltonian of our artificial molecule is
\be\label{H1}
\hat{H}=-\frac{\Delta_1}{2}-\frac{\Delta_2}{2}
-\sum_{j,i=1}^2\frac{1}{\sqrt{h_j^2+(x_i-(-1)^j a)^2 +y_i^2}} +
\frac{1}{r_{12}}+\frac{1}{\sqrt{R^2+(h_2-h_1)^2}}
\ee
Hereafter we use effective atomic units (a.u.) which means that all 
distances are measured in units of the effective Bohr radius 
$a_B=\hbar^2\epsilon/m^*e^2$, and energies in units of 
$m^*e^4/\hbar^2\epsilon^2$, where $m^*$ is the effective
 mass of the electron, and $\epsilon$ is the dielectric constant. 
The projection of the distance between impurities  onto the plane of electron gas
is $R\equiv 2a$ and the $x$-coordinates of the impurities 
are $\pm a$.   

 Three parameters, which determine the behavior of the system, are  $h_1,\ h_2,$ and
$R$. Of course, there is no analytical solution for $J(h_1,h_2,R)$ for 
the whole range of  parameters. In the present work we investigate the
different analytical limits of the Hamiltonian (\ref{H1}) and present an
interpolated formula, which connects these limits.

 If $h_1\sim h_2\sim h$ the method of the calculations is primarily determined 
 by the ratio $h/R$. This can be understood by looking at the single-particle
unperturbed potential (see Fig. 2).
\be\label{V1}
V_1(x,y)=
-\frac{1}{\sqrt{h_1^2+(x+ a)^2 +y^2}} 
-\frac{1}{\sqrt{h_2^2+(x- a)^2 +y^2}}. 
\ee

\begin{figure}\label{fig2}
\epsfxsize=14 truecm
\epsfysize=7 truecm
\centerline{\epsffile{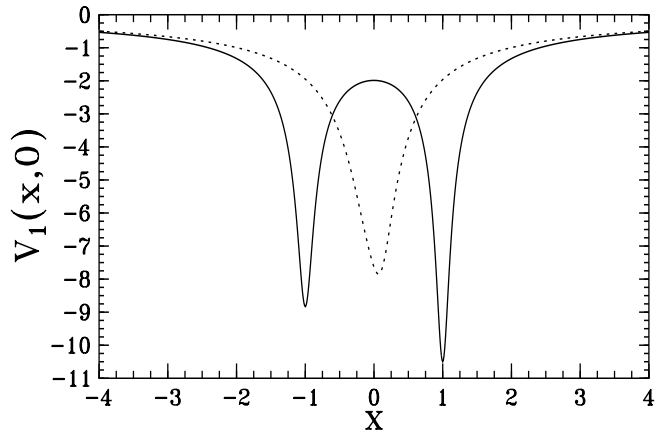}}
\caption{The cross section $y=0$ of the single-particle potential $V_1(x,y)$.
The dotted line is the Helium-like atom ($h_1,h_2>R$), and
the solid line is the Hydrogen-like molecule ($h_1,h_2<R$).}
\end{figure}
Function $V_1(x,y)$ has either  one or two minima. The transition from one case to another happens
when a minimum near the origin changes for a  maximum.  
For the case of equal heights $h_1=h_2=h$ the function 
$V_1(x,y)$ has one minimum if  $h/R\geq 1/\sqrt{2}$.

 The case $h_1,h_2 < R$ corresponds to a $2D$ Hydrogen-like molecule,
when there are two well-separated minima.
In this case the splitting is exponentially small and determined by
the degree of overlapping of wave-functions in the classically forbidden
under-barrier region (especially at $r_1,r_2\approx 0$).

 The opposite case $h_1,h_2>R$ corresponds to a $2D$ Helium-like atom when 
only one minimum exists, and the interaction term $1/r_{12}$
should be taken into account from the very beginning. 

  If the difference between $h_1$ and $h_2$ is large enough then
the electrons will spend  most of their time in the potential well
of the impurity closest to the plane. The potential $V_1(x,y)$
 has one minimum in this case. Therefore this case corresponds
to a perturbed ``Helium'' atom  (or perturbed ion $H^{-}$).

 The  paper is organized as follows.  
In Section 2 we consider the case of a hydrogen-like molecule ($R>2h$).
By making use of
semiclassical methods we obtain analytical results for a pure 
two-dimensional hydrogen molecule ($h_1,\ h_2=0,\ R\gg 1$)\cite{pon}. The perturbation
theory results  ($h_{1,2}\ll 1<R$) and the  saddle point method ($1< h_{1,2}
\ll R$)
represent other analytical limits in this case. The details of the calculations are given in appendices \ref{ApB},\ref{ApC}.

In Section 3 we discuss a helium-like atom ($h_{1,2}>R$). We start from the result
 for pure two-dimensional helium\cite{pon}
($R=h_{1,2}=0$) and  calculate the  perturbation theory corrections. The
 quadratic approximation is applied to the limit $h_{1,2}>R>1$. 
Section 4 is devoted to the interpolated formula, which
connects the two different limits for the distance $R$ at fixed vertical coordinates of impurities 
 $h_1,h_2$. Our results are summarized in the Conclusion.

\section{Hydrogen-like molecule}
\subsection{General formula, $R\gg 1$.}
 The exchange constant $J_{ik}$ is defined as one-half of the energy
difference between the lowest two-particle energies for total spins $S=1$ 
and $S=0$: 
\begin{eqnarray}
2J &=&(E_g^{S=1}-E_g^{S=0})\  \equiv (E_A-E_S),\quad
\mbox{ where }\label{J}\\
\hat{H}\Psi_S &=& E_S\Psi_S,\label{eq1}\\
\hat{H}\Psi_A &=& E_A\Psi_A.\label{eq2}
\end{eqnarray}
Here  $\Psi_S$ and $\Psi_A$ denote the lowest symmetrical and antisymmetrical 
two-particle coordinate wave functions of the singlet and triplet states 
respectively. 
When $h_{1,2}<R$ and $R\gg 1$ the most appropriate method to 
determine  the term splitting due to the spin-spin interaction
is the Gor'kov-Pitaevskii method \cite{Gorkov,Smirnov,Flam98}.
 We apply it here to
the two-dimensional electron system.
In accordance with this approach we reduce the expression for $J$ to a 
surface integral over a hyperplane in the $4$-dimensional coordinate space.
To this end, we multiply  Eq.~(\ref{eq1}) by $\Psi_A$ and  
Eq.~(\ref{eq2}) by $\Psi_S$,
take the difference between the results and calculate the integral over
{\em a part of the volume} in the four dimensional configuration  space of 
the electrons.
We choose the integration volume as a region $x_1\leq x_2$. 
This is a four-dimensional volume to the left of the hyperplane $\Sigma$ determined 
by the equation  $x_1=x_2$.
Using the Hamiltonian (\ref{H1}) and Gauss' theorem we obtain
 \be\label{split1}
(E_S-E_A)\int\!\!\int_{\Omega}\Psi_A\Psi_S\, d{\bf r_1}\,
d{\bf r_2}=
\oint_{\sigma}(\Psi_S\nabla \Psi_A -\Psi_A\nabla\Psi_S)\,{\bf d\sigma},
\ee
where $\sigma$ is a closed hypersurface, and $\Omega$ is the volume bounded by
$\sigma$.

Now we introduce a combinations of the functions 
$\Psi_{1,2}=1/\sqrt{2}
(\Psi_S\pm\Psi_A)$.
If the phases of $\Psi_{S,A}$ are properly chosen, the function    
$\Psi_1$ will be large only when electron 1 is near the left minimum and
electron 2 is near the right minimum. 
The function $\Psi_2$, obtained by permuting 
${\bf r_1}$ with ${\bf r_2}$, is localized almost entirely on the ``far'' side.
A simple calculation gives
$$\int\!\!\int_{\Omega}\Psi_S\Psi_A\, d{\bf r_1}\,
d{\bf r_2}=\frac{1}{2}\int\!\!\int_{\Omega}(\Psi_1^2-\Psi_2^2)
\, d{\bf r_1}\,d{\bf r_2}\approx 1/2.
$$
Substituting the wave functions $\Psi_{1,2}$ into Eq.~(\ref{split1})
we finally obtain that $(E_S-E_A)$ can be expressed as the integral  over the  hyperplane
$\Sigma$.
\be\label{JvsInt}
2J=-2\int\left [\Psi_{2}\frac{\partial\Psi_1}{\partial x_1}-
\Psi_{1}\frac{\partial\Psi_2}{\partial x_1}
\right]_{x_1=x_2}~dx_2~dy_1~dy_2.
\ee

 Eq.~(\ref{JvsInt}) shows that the main contribution to the
exchange constant is given by the region where the electrons are close to
each other. Indeed, the $x$ coordinates of both electrons coincide 
($x_1=x_2$), however, the $y$ coordinates may
be different. In this case there are strong correlations between the 
positions of the electrons due to Coulomb repulsion. This means that we
should go beyond the approximation where the two-particle wave function 
of the electrons is represented as a product of single-particle 
wave functions. 

 The details of the calculation of the wave functions $\Psi_{1,2}$ and 
the surface
integral (\ref{JvsInt}) are presented in the
appendices \ref{ApB},\ref{ApC}.
The final result is:
\be\label{Jalbet1}
2J(\alpha,\beta,R)=
R^{\frac{2}{\alpha}+\frac{2}{\beta}-\frac{1}{\mu}}e^{-\mu R}
\left[D\left(\alpha,\beta,R\right)+D\left(\beta,\alpha,R\right)\right],
\ee
where the function $D\left(\alpha,\beta,R\right)$ is determined by 
Eq. (\ref{Dabr}).
Here 
$$\frac{\alpha^2}{2}=-E(h_1),\ \frac{\beta^2}{2}=-E(h_2)$$
are ``ionization energies'' in the potential   $1/\sqrt{r^2+h^2}$,
and  $\mu=\alpha+\beta$.

\subsection{Two-dimensional hydrogen molecule and
 perturbation theory for $h_1,\ h_2\ll 1$}

In the case $\alpha=\beta$ the function $D(\alpha,\alpha,R)$ is 
independent of $R$:
\be\label{D0a1}
D_0(\alpha)\equiv 2\,D\left(\alpha,\alpha,R\right)=
8\sqrt{\pi}A_{\alpha}^4\left(\frac{1}{4\alpha}\right)^{1/2\alpha}
\Gamma\left(\frac{\alpha+1}{2\alpha}\right)
\int_0^1 \exp\left(-t/\alpha\right)
(2-t)^{1/2\alpha}t^{3/2\alpha}\,dt.
\ee
Therefore for the case $h_1=h_2$
\be
\label{Jf1}
2J(\alpha,\alpha,R)=D_0(\alpha)R^{7/2\alpha}\exp(-2\alpha R)
\ee
Here $A_{\alpha}$ is the normalization constant for the  single particle
wave function in the potential $1/\sqrt{r^2+h^2}$ (see Appendix \ref{ApB}).

The case $h=0$ corresponds to the two-dimensional hydrogen molecule.
In accordance with Appendix \ref{ApB} we substitute $\alpha=2$ and 
$A_{\alpha}=4/\sqrt{2\pi}$ in Eq. (\ref{Jf1}) and we obtain  
\be\label{H_2t}
2J(2,2,R)=30.413\, R^{7/4}\exp(-4R)
\ee
 
We can carry out further analytical estimates for small 
$h_{1,2}$.
When $h$ is much smaller than the Bohr radius we take $A_{\alpha}$ from 
(\ref{Anorm}), put it into  Eq. (\ref{D0a1}) and obtain
\be\label{D0af}
D_0(\alpha)=
\frac{2^{3/\alpha+1}\alpha^{4+7/2\alpha}}{\pi^{3/2}}
\frac{\Gamma\left(1/2+1/2\alpha\right)}{\Gamma\left(2/\alpha\right)^2}
\int_0^1\exp\left(-t/\alpha\right)\left(2-t\right)^{1/2\alpha}t^{3/2\alpha}
\,dt
\ee
Function $D_0(\alpha)$ is a regular function near the point $\alpha=2$
(see Fig. 3). It can be expanded in a Taylor series:
\be
D_0(\alpha)\approx 30.413\left(1+1.665(\alpha-2)+0.456(\alpha-2)^2+\ldots
\right)
\ee
\begin{figure}\label{fig3}
\centerline{\epsffile{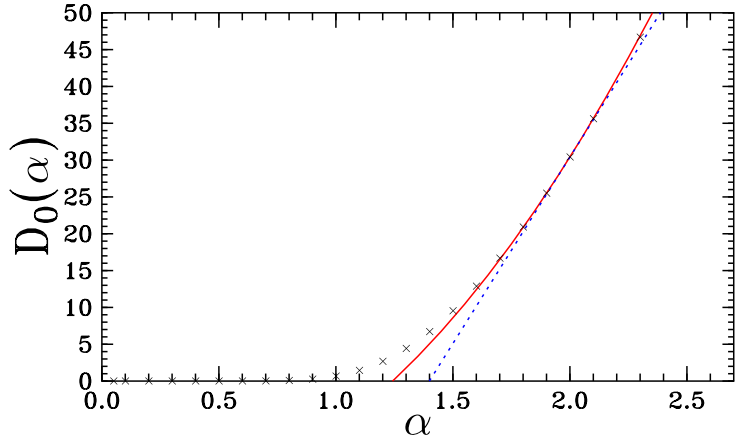}}
\caption{Exact (crosses), linear (dashed line) and quadratic (solid
line) approximations for function $D_0(\alpha)$. 
} 
\end{figure}
 Taking into consideration that (see Appendix \ref{ApB})
$$\alpha\approx 2-\Delta_E/2$$
with 
$$\Delta_E=16h\left(1+2h\ln(2h)\right),$$ 
finally we obtain
\be\label{Jfa}
J(\alpha,\alpha,R)\approx J(2,2,R)\left[1-0.833\Delta_E\right]
R^{\frac{7}{16}\Delta_E}\exp(\Delta_E R)\approx
J(2,2,R)\left[1+\Delta_E(-0.833+R+\frac{7}{16}\ln(R))\right].
\ee
Thus, the splitting increases with $h$. Indeed, when the
impurity centers go out of the plane the interaction between them and
the electrons becomes smaller. Therefore, 
radii of bound single electron wave functions and overlapping increase.
 
When $h_1\neq h_2$ and $h_1,h_2\ll 1$ the formula for $J$ should
be rewritten in the form:
\be\label{Jfab}
J(\alpha,\beta,R)\approx
J(2,2,R)\left[1+\Delta_E(-d_1+R+7/16)+\delta_E d_2\right],
\ee
where 
$$\Delta_E,\delta_E=(\Delta_{E_1}\pm\Delta_{E_2})/2$$
and $d_1,d_2$ are some coefficients which can be extracted from
the expansion $D(\alpha,\beta,R)$ for small $h_{1,2}$.

\subsection{$1\ll h\ll R$. Saddle point method.} 
At $h\gg 1$, one can easily find that  $E(h)\approx -1/h$ (See also Appendix \ref{ApB}). Then $\alpha
\approx
\sqrt{2}/\sqrt{h}\ll 1$.
and
the integral in $D_0(\alpha)$ as well as the Gamma-function can be 
estimated by the saddle-point method:
\bea
\int_0^1e^{-t\lambda}(2-t)^{\lambda/2}t^{3\lambda/2}\,dt &\approx&
\frac{e^{-\lambda/2}}{2}\sqrt{\frac{\pi}{\lambda}},\nonumber\\
\Gamma(\lambda/2+1/2)&\approx& \sqrt{\pi(\lambda-1)}\left(
\frac{\lambda-1}{2}\right)^{\frac{\lambda-1}{2}}
\exp\left(-\frac{\lambda-1}{2}\right),\nonumber\\
\eea
where $\lambda\equiv 1/\alpha$.
 Thus,
\be
D_0(\alpha)\approx \frac{4}{\pi}\frac{1}{h^3}
\frac{e^{-\lambda}}{\sqrt{\lambda}}\left[\frac{\lambda}{4}\right]^{\lambda/2}
\sqrt{2\pi}\left[\frac{\lambda}{2}\right]^{\lambda/2}e^{-\lambda/2}=
\sqrt{\frac{32\alpha}{\pi}}\frac{1}{h^3}
\left[\alpha^2 8e^3\right]^{-1/(2\alpha)},
\ee
and
\be\label{Jspaa}
2J(\alpha,\alpha,R)=
\sqrt{\frac{32\alpha}{\pi}}\frac{1}{h^3}
\left[\alpha^2 8e^3\right]^{-1/(2 \alpha)}
R^{7/(2\alpha)}\exp(-2\alpha R),
\qquad h\gg 1,R/h\gg 1.
\ee

\section{Helium-like atom}
\subsection{Pure $2D$ Helium atom.}

 When two impurities are close to each other ($R\lesssim \sqrt{2}h$) there is
only one minimum for the single-particle potential $V_1(x,y)$. Hence,
the semiclassical expression (\ref{JvsInt}) for the exchange constant 
is not applicable anymore and we have to exploit other methods.

 For the case $h_{1,2}=0,\ R\approx 0$ we use a variational approach
for the two-dimensional helium atom. 
 We consider impurities as the ``nucleus''  with a charge $Z=2$ and the
trial functions are
\begin{eqnarray}
\Psi_S&=&e^{-\alpha s}\cosh(\beta t),\nonumber\\
\Psi_A&=&e^{-\alpha s}\left[\alpha t\cosh(\beta t)+
\left(\alpha s-2\right)\sinh(\beta t)\right],\nonumber\\
s,t&=& r_1\pm r_2.
\end{eqnarray}
Here $\alpha$ and $\beta$ are free parameters, which have to be
determined from the variational principle.
The details of the calculations were presented in\cite{pon}.
We obtained the following results for the energies:
\begin{eqnarray}\label{2dhel}
E(^{1}S)&=&-11.635\ \mbox{a.u.}\nonumber\\
E(^{3}S)&=& -8.193\ \mbox{a.u.}.
\end{eqnarray}
 Thus, in this case
\begin{equation}\label{2dhelf}
2J=3.567\, (\pm 1\%) \mbox{a.u.}
\end{equation}
\subsection{Small corrections due to finite size of the ``nucleus''}

When the distance between impurities (or distance from impurities to
 the plane) is small in comparison with the Bohr radius
we can use first-order perturbation theory to estimate the level's energy 
shift:
\be
\Delta E =\int\Psi^2(r_1,r_2)\sum_{i=1}^2\Delta V(\vec{r}_i)\, dV_1\,dV_2.
\ee
But the integral over all variables except one is equal to
\be
\int\Psi^2(r_1,r_2)\,dV_2=\frac{1}{2}\rho(r_1),
\ee
where $\rho(r)$ is the electron density (normalized to two - the number
of  particles).
Therefore,
\be\label{PT1O}
\Delta E =\int\rho(r)\Delta V(\vec{r}_1)\, dV\approx
\rho(0)\int_0^{2\pi}\int_0^{\infty}\exp(-2\alpha r)\Delta V(\vec{r}_1)\,
r\,dr\,d\phi.
\ee
Here we take into consideration the fact that the main contribution
from density gives the term which is simply proportional to the exponent
(for the $S=0$ state $\alpha_0=2Z_0$ and for the $S=1$ state 
$\alpha_1=Z_1+Z_2/3$).

For the case $h_1=h_2=0$, $R\ll 1$
the difference between the point ``Helium'' nucleus and the real potentials
for the single electron is
\be\label{DV}
\Delta V(\vec{r})=\frac{2}{r}-\frac{1}{\sqrt{r^2+a^2-2ra\cos(\phi)}}
-\frac{1}{\sqrt{r^2+a^2+2ra\cos(\phi)}}
\ee

The double integral for the last two terms of  (\ref{DV}) can be 
reduced to the single integral
\begin{equation}
8a\left[\int_0^1 K(x)\exp(-2 a\alpha x) x\,dx+
\int_1^{\infty} K(1/x)\exp(-2 a\alpha x)\,dx
\right].
\end{equation}
Here $K(x)$ is a complete elliptic integral of the first kind.

For an estimate of the first integral  we can use the
fact that $\exp(-2 a\alpha x)\approx 1$ in the region $r<a$ and
in the second integral we put 
$$K(1/x)\sim \frac{\pi}{2}+\ln(\frac{x}{\sqrt{x^2-1}}).$$
 Thus
\be
\Delta E=8a\,\rho(0)\left(1+\ln(2)\right),
\ee
 
and hence, the exchange constant is
\be\label{JHeR}
2J(0,0,R)=2J(0,0,0)-4\left(1+\ln(2)\right)\Delta\rho(0)R=3.567-30.4 R.
\ee
Here
\be
\Delta\rho(0)=\rho_0(0)-\rho_1(0)=
2\frac{16}{2\pi}
\left[Z_0^2-\frac{1}{2}\left(Z_1^2+\frac{Z_2^2}{27}\frac{8Z_2^2}
{9Z_1^2-2Z_1Z_2+Z_2^2}\right)\right]\approx 4.49.
\ee

 For the case $R=0$ and $h_1,h_2\ll 1$
\bea
\Delta E &=&2\pi\rho(0)\int_0^{\infty}\exp(-2\alpha r)
\left(\frac{2}{r}-\frac{1}{\sqrt{h_1^2+r^2}}-\frac{1}{\sqrt{h_2^2+r^2}}
\right)r\,dr \approx\nonumber\\
&\approx & 2\pi\rho(0)\left[h_1+h_2+\alpha h_1^2\ln(\alpha h_1)
+\alpha h_2^2\ln(\alpha h_2)\right]
\eea
Therefore,
\be\label{JHeh}
J(h_1,h_2,0)\approx 2J(0,0,0)-
2\pi\left(h_1+h_2\right)\Delta\rho(0)=3.567-28.2(h_1+h_2).
\ee
It is worthwhile to stress that the convergence 
region of the  perturbation theory is very small ($R< 0.1 a_B$) because of the large numerical factors
in the second terms of Eqs. (\ref{JHeR}) and (\ref{JHeh}) . 
\subsection{Case $R\ll h^{3/4}$, $h\gg 1$. Quadratic approximation}\label{hggR}

If the impurities are far enough from the plane $h_1\sim h_2=h\gg 1$ and 
$\sqrt{2}h>R$, the
single-particle potential (\ref{V1}) has an oscillatory shape near
its single minimum.
 Therefore, taking into consideration its Taylor expansion near zero the
 potential in two-electron hamiltonian (\ref{H1}) has the following form:
\bea\label{Uh1h2}
U(x_1,x_2,y_1,y_2)&=& -\frac{4}{\sqrt{h^2+a^2}}+
\frac{h^2-2a^2}{\left(h^2+a^2\right)^{5/2}}\left[
(x_1^2+x_2^2+y_1^2+y_2^2\right]+\nonumber\\ 
&+&\frac{3a^2}{\left(h^2+a^2\right)^{5/2}}\left[(y_1^2+y_2^2\right]+
\frac{1}{\sqrt{(x_1-x_2)^2+(y_1-y_2)^2}}
+O\left(\left(h^2+a^2\right)^{-5/2}\right).
\eea

Introducing the notations:
$$\omega_{+}^2=2\frac{h^2-2a^2}{\left(h^2+a^2\right)^{5/2}},\
\omega_{-}^2=\frac{6a^2}{\left(h^2+a^2\right)^{5/2}}$$
and substituting the variables
\begin{eqnarray}\label{chv}
\xi,\eta &=&\frac{ x_1 \pm  x_2}{\sqrt{2}},\\
u,v&=&\frac{y_1\pm y_2}{\sqrt{2}}\nonumber
\end{eqnarray}
we separate center-mass ($\xi,u$) and relative motion ($\eta,v$):
\bea\label{cmrm}
\hat{H}_{cmm}&=& -\frac{\Delta_{\xi,u}}{2}
+\frac{\omega_{+}^2}{2}\left[\xi^2+u^2\right]+
\frac{\omega_{-}^2}{2}u^2\nonumber\\
\hat{H}_{rm}&=& -\frac{\Delta_{\eta,v}}{2}
+\frac{\omega_{+}^2}{2}\left[(\eta^2+v^2\right]+
\frac{1}{\sqrt{2}\sqrt{\eta^2+v^2}}
+\frac{\omega_{-}^2}{2}v^2=\nonumber\\
&=&-\frac{\Delta_{r,\varphi}}{2}
+\frac{\omega_{+}^2}{2}r^2+\frac{1}{\sqrt{2}r}
+\frac{\omega_{-}^2}{2}r^2\sin(\varphi)^2
\label{Hrm}
\eea

The energy splitting between the singlet and the triplet states is 
determined by the relative motion part of the Hamiltonian.
If we neglect  by the last "perturbed" term in (\ref{Hrm}),
the potential for the relative motion is independent of $\varphi$. Therefore
$$\Psi(\eta,v)=\phi(r)\exp(i m \varphi)/\sqrt{2\pi}.$$

 The state with  $S=0$ corresponds to an even function of coordinates and the state with 
$S=1$ corresponds to an odd function. Parity is determines by
the two-dimensional angular momentum $m$.
$$\vec{r}\rightarrow -\vec{r}\Rightarrow r\rightarrow r,\ 
\varphi\rightarrow \pi+\varphi.$$
Thus, symmetric ($S=0$) wave functions correspond to 
even angular momenta ($|m|=0,2,4...$) and antisymmetric wave functions
correspond to odd angular momenta ($|m|=1,3,5...$)

 The exchange constant is $2J=E_{01}-E_{00}$, where $E_{n_r,|m|}$
are the energy levels for the potential
$$U(r)=  \frac{\omega_{+}^2}{2}r^2+\frac{1}{\sqrt{2}r}.$$
This potential can be approximated by an oscillatory one after expanding 
around its equilibrium position $r_0=2^{-1/6}/\omega_{+}^{2/3}$:
$$U(r)\approx U(r_0)+\frac{3\omega_{+}^2}{2}\left(r-r_0\right)^2$$
 Thus, in this limit we obtain
\be\label{Jhgg1}
2J\approx\sqrt{3}\omega_{+}=
\left(9\frac{[2h^2-R^2]^2}{[h^2+R^2/4]^5}\right)^{1/4},
\quad R\ll h.
\ee
Taking into account that first-order corrections must be small in comparison 
with the energy splitting we obtain more accurate condition for the
applicability  of Eq. (\ref{Jhgg1}):
\begin{equation}\label{conJhgg1}
\frac{<\Delta \hat{H}_{rm}^{(1)}>}{2J}\sim 
\frac{\omega_{-}^2r_0^2}{\omega_{+}}\ll 1. 
\end{equation}
It is equivalent to the condition  $R\ll h^{3/4}$.
\section{Interpolation}
Let us consider equal heights of the impurities above the plane: $h_1=h_2=h$.
In the asymptotic region, for large distances between the impurities, we have
\be\label{Jf1i}
2J(h,R)=D_0(\alpha)R^{7/2\alpha}\exp(-2\alpha R),\qquad R\gg h.
\ee
Here $D_0(\alpha)$ is determined by Eq. (\ref{D0a1}) and $-\left[\alpha(h)\right]^2/2$ is 
the single-particle ground state energy in the potential $-(r^2+h^2)^{-1/2}$ 
(see Appendix \ref{ApB}).
In the opposite helium-like limit we have
\be\label{hellike}
2J(h,0)=\left\{ 
  \begin{array}{ll}
-3.567,& h=0\\
\sqrt{\frac{6}{h^3}}, & 1\lesssim h.
\end{array} \right.
\ee
Let us now make a plausible assumption that 
for small  distances between the impurities ($R\lesssim h$) the behavior of 
the exchange constant is
\be\label{insm}
2J(h,R)\approx 2J(h,0)\exp\left(-\gamma R -\gamma_2 R^2\right),
\ee  
where $\gamma=\gamma(h)$ is a free parameter, and $\gamma_2=\gamma_2(\gamma)$.
Such form of the functional dependence follows from the fact that 
the most accurate numerical calculations for the
three-dimensional hydrogen molecule\cite{wol} are  fitted  accurately
at small distances by the  formula $2J(0)\exp(-\gamma_1R-\gamma_2R^2)$
with just two independent  parameters $\gamma_1,\ \gamma_2$\cite{fit}.
Let us  match the second derivative of
$\ln\left(2J(h,R)\right)$.
In the two regions it must have the following behavior 
[See Eqs. (\ref{Jf1i},\ref{insm})] 
\begin{equation}\label{lnlim}
\frac{\partial^2\ln(2J)}{\partial R^2}\approx\left\{
\begin{array}{ll}
-2\gamma_2, & R\leq h\\
-\frac{7}{2\alpha R^2}, & R\gg h.
\end{array}\right.
\end{equation}
 The simplest formula that satisfies  both conditions is
\begin{equation}
\frac{\partial^2\ln(J)}{\partial R^2}=-\frac{2\gamma_2}
{1+4/7\gamma_2\alpha R^2}.
\end{equation}
 After  integrating  over $R$ twice we obtain
\begin{eqnarray}\label{int3D}
\ln(2J) &=&C-\gamma R-
\frac{7}{2\alpha}AR\arctan(AR)+\frac{7}{4\alpha}\ln(1+A^2R^2).
\end{eqnarray}
where $C$ and $\gamma$ are integration constants and $A=\sqrt{4\alpha\gamma_2/7}$.
It is obvious that we have to put $C=\ln\left(2J(h,0)\right)$ in order to fulfill
(\ref{insm}). On the other hand to guarantee the correct asymptotic exponent  
of Eq. (\ref{Jf1i}) we have to impose a constrain
\be\label{constr}
A=\frac{4\alpha(2\alpha-\gamma)}{7\pi}.
\ee
Finally,  the interpolated formula has the form
\begin{equation}\label{Jinter}
2J(h,R)=2J(h,0)\left[1+A^2R^2\right]^{7/4\alpha}\exp\left[-R\left(\gamma
+\frac{2(2\alpha-\gamma)}{\pi}\arctan(AR)\right)\right].
\end{equation}

Two parameters in this formula,  $J(h,0)$ and $\gamma(h)$, are not defined yet. Additional
numerical calculations  are needed to determine them. 
The parameter $2J(h,0)$ decreases from  3.567 at 
$h=0$ to $\sqrt{6/h^3}$ for
$h\gg 1$ [see Eq. (\ref{hellike})]. For intermediate values of $h$ 
it  can be determined by applying a variational approach similar to that of 
Ref. \cite{pon},  to the helium-like atom. We can do a rough estimate of $J(h,0)$ using an 
interpolation between these two limits:
\begin{equation}
2J(h,0)=\frac{3.567}{1+\frac{3.567h^{3/2}}{\sqrt{6}}}
\end{equation}

  The parameter $\gamma(h)$ can be estimated by considering small and large $h$. In our previous
paper\cite{pon} we have found a reliable interpolated formula for splitting in the case $h=0$. It reads
\begin{equation}
 2J(0,R) = 3.567\left(1+1.81R^2\right)^{7/8}\exp
\left(-0.3 R-2.355R\arctan\left(1.346R\right)\right).
\label{int2D}
\end{equation}  
 To get this equation from Eqs. (\ref{Jinter}),(\ref{constr}) at $h\rightarrow 0$ one should put
$\gamma(0)=0.3$. We can argue that $\gamma(h)$ should decrease with increasing $h$.
Indeed, Eq. (\ref{Jhgg1}) shows that function $J(h,R)$ is an algebraic rather than exponential function
of $R$ at
$R\ll h^{3/4}$.  To satisfy this condition one should assume that  $\gamma(h)\sim h^{-3/4} $ at $h\gg
1$. Then it follows from Eq. (\ref{constr}) that $A=8\alpha ^2/7\pi$ at $h\gg 1$, where $\alpha$ is
determined by Eq. (\ref{oscen}). Thus $A\sim 1/h$ at $h\gg 1$.

 The function $\gamma(h)$ can be chosen, for example, in the following form
$$\gamma(h)=\frac{0.3}{1+h^{3/4}}.$$ 
It takes into consideration the main features. We would like to stress that
there is practically no dependence  on 
$\gamma$ in the exchange constant (\ref{Jinter}), since it is small even for $h=0$ and
 it disappears completely due to cancellation of two terms in Eq. (\ref{Jinter}) at large $R$.

  The results are summarized in Fig. 4, where a semi-logarithmic plot
of the exchange constant vs. $R$ for three different heights is presented.
The solid line shows $2J(0,R)$ for the pure two-dimensional hydrogen molecule.
The curve $h=0.8$ corresponds to the atomic parameter $\alpha\approx 1$ and,
consequently, it has a two times smaller
asymptotic exponent in the exchange constant. Of course, this case corresponds
to the intermediate values of the parameter $h$. Therefore, an approximate realistic
magnitude of $2J(h,0)$ was chosen.
The last curve shows that in the limit of large $h$ the
behavior of the splitting energy is flat, and almost independent of $R$ for
$R\ll h^{3/4}$, in accordance with Eq. (\ref{Jhgg1}). The value of $\alpha$ is taken
from the oscillator limit (\ref{oscen}) and $2J(h,0)$ corresponds to the
lower formula in (\ref{hellike}).
\begin{figure}\label{fig4}
\epsfxsize=14 truecm
\epsfysize=9 truecm
\centerline{\epsffile{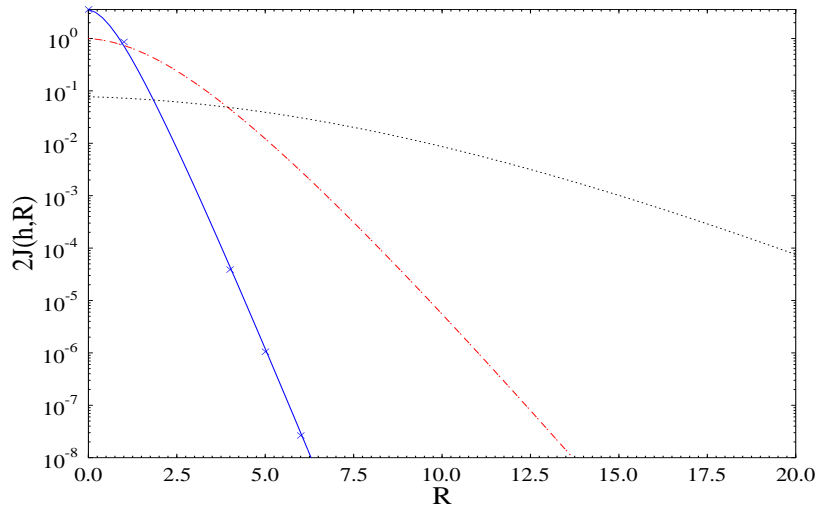}}
\caption{The exchange constant as a function of $R$ and
$h$. The curves for three different heights are presented.
The crosses correspond to the numerical data for $R=0$ and $R=1$ and asymptotic
form (\ref{H_2t}) for the pure two-dimensional case ($h=0$). The solid line
is the interpolated formula (\ref{int2D}). Dot-dashed and dotted lines
present the interpolated formula (\ref{Jinter}) with $h=0.8$ and 10 respectively.
}
\end{figure}

\section{Conclusion}
 We have calculated analytically different  limits of the 
singlet-triplet energy splitting in the two-electron Hamiltonian (\ref{H1}) in the
following cases.
\begin{enumerate}
\item Hydrogen-like molecule.
  \begin{enumerate}
   \item $h_1,h_2 =0,  R> 1$, which corresponds to a $2D$ Hydrogen molecule, 
                Eq. (\ref{H_2t}).
   \item $h_{1,2}\ll 1,\ R>1$, a Hydrogen-like molecule, where 
 perturbation theory is applicable,
                  Eqs. (\ref{Jalbet1},\ref{Jf1},\ref{Jfa},\ref{Jfab}).
   \item $1\ll h_1, h_2 \ll R$, saddle point method,
       (Eq. (\ref{Jspaa})).
 \end{enumerate}
\item Helium-like atom.
  \begin{enumerate}
 \item $R=h_1=h_2=0$ corresponds to the $2D$ Helium atom, Eq. (\ref{2dhelf}).
 \item $R\ll 1, h_1=h_2=0$, $2D$ Helium atom with 
perturbation theory corrections, Eq. (\ref{JHeR}).
 \item $R=0, h_1,h_2\ll 1$, $2D$ Helium atom with
 perturbation  theory corrections, Eq. (\ref{JHeh}). 
 \item $h_1,h_2>R>1$, quadratic approximation for the Helium-like atom,
                                                   Eq. (\ref{Jhgg1}).
 \end{enumerate}
\end{enumerate}
   
  Finally, we have presented the interpolated formula (\ref{Jinter}) 
connecting the limits 1 and 2. This formula can be used for calculations
of the ground state in spin glasses and semiconductors based on the Heisenberg
Hamiltonian. We have shown that  the
$R$-dependence of the splitting  becomes weaker with increasing $h_1,h_2$. 

\acknowledgments
This work is supported by the Australian Research
Council and by the Seed Grant of the University of Utah.  I. Ponomarev acknowledge 
fruitful discussions with G. Gribakin. 
\appendix
\section{Eigenvalues and eigenfunctions in $2D$ central potentials} 
\label{ApB}

The Schr\"{o}dinger equation for a radial wave function in a central field is 
\be\label{hamH}
\left \{\frac{\partial^2}{\partial r^2} +
\frac{1}{r}\frac{\partial}{\partial r} -
\frac{m^2}{r^2} -2U(r)+2E
\right\}R(r)=0.
\ee
 For the Coulomb potential, $U(r)=-Z/r$, the eigenvalues and eigenfunctions are
well-known:
\bea
E_n&=&-\frac{Z^2}{2\,(n-1/2)^2},\\
n&=& 1,2,3...\nonumber\\
m&=&-n+1,-n+2,...n-1\nonumber\\
\varphi_{nm}(\vec{r})&=&\frac{Ze^{im\phi}}{\sqrt{2\pi}}
\sqrt{\frac{16\,(n-|m|-1)!}{(2n-1)^3\left [(n+|m|-1)!\right]^3}}
\left [\frac{4rZ}{2n-1}\right ]^{|m|}\exp\left (-\frac{2rZ}{2n-1}\right )
L_{n+|m|-1}^{2|m|}\left (\frac{4rZ}{2n-1}\right ),
\label{wfHyd}
\eea
where $L_{l}^{k}(\rho)$ is the generalized Laguerre polynomials, and the level
degeneracy  is $g(n)=2n-1$.

The most important for applications here $1s$ and $2s$ states have the following form.
\bea\label{PsiH0}
\varphi_{10}(r)&=&\frac{4Z}{\sqrt{2\pi}}e^{-2rZ}\nonumber\\
\varphi_{20}(r)&=&\frac{4Z}{3\sqrt{6\pi}}e^{-2rZ/3}\left [1-\frac{4rZ}{3}
\right ]\nonumber\\
\eea
Thus, for the ground state with $Z=+1$
\begin{eqnarray}
\alpha &=&2\nonumber\\
A_{\alpha} &=& \frac{4}{\sqrt{2\pi}}
\end{eqnarray}

Let us consider now the potential 
\begin{equation}
U(r)=-1/\sqrt{r^2+h^2}.
\end{equation}
 Far away from the ``atomic residue''  $U(r)\sim -1/r$ and
the  wave function for the $s$-state obeys the Schr\"{o}dinger
equation
$$-\frac{\Delta}{2}\varphi-\frac{1}{r}\varphi=-\frac{\alpha^2}{2}\varphi$$
with the solution
\be\label{assymH}
\varphi(r)=A_{\alpha}r^{1/\alpha-1/2}e^{-\alpha r}
\ee
up to $\varphi/r^2$ accuracy.
Here $A_{\alpha}$ and $\alpha$ are atomic parameters. Their magnitudes
are determined by the behavior of the electron inside the ``atom''.

 The results of the numerical calculations for the normalization constant and
the eigenvalues are presented in  Figure~5.
\begin{figure}\label{fig5}
\centerline{\epsffile{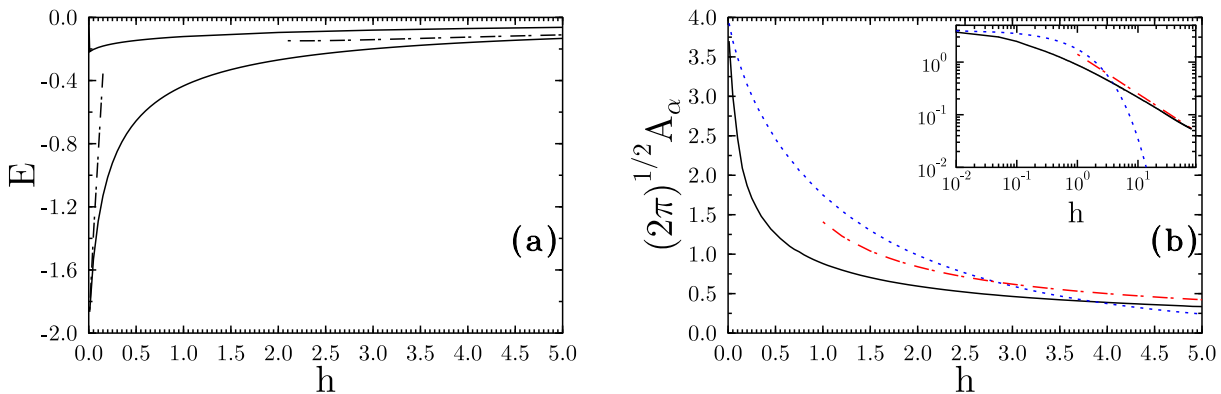}}
\caption{ (a) Ground and first energies as functions of the parameter $h$.
Solid lines are numerical results. The dot-dashed lines correspond to perturbation theory for small 
$h$ and an
asymptotic oscillatory energy $-h^{-1}+h^{-3/2}$ for $h\gg 1$.
(b) The constant $A_{\alpha}$ for the ground state. Numerical results (solid line)
and analytical estimates for both regimes are shown. Dotted line corresponds to Eq. 
(\protect\ref{Anorm}) and dot-dashed line is the quadratic approximation (\protect\ref{next}).
The inset is a log-log scale plot.}
\end{figure}
Simple analytical estimates for the cases of small and large $h$
 for ground state energy and the asymptotic
coefficient $A_{\alpha}$ can be made.

 When $h\ll 1$ the potential is only slightly different from the Coulomb 
potential
and the contribution to the energy can be obtained using 
perturbation theory:
\be\label{pe}
\Delta_E=\langle 0|V|0\rangle =16\int_0^{\infty}
\left(\frac{1}{r}-\frac{1}{\sqrt{h^2+r^2}}\right)\exp(-4r)r\,dr
\approx 16h\left(1+2h\ln(2h)\right)
\ee
Note, that due to the big factor in front of $h$ in (\ref{pe})
the convergence radius for the perturbation series is small enough. 

 Supposing that the asymptotic form (\ref{assymH}) 
of the wave function is valid for all values of $r$, we obtain the following
 simple estimate for $A_{\alpha}$ 
\be\label{Anorm}
2\pi A_{\alpha}^2\int_0^{\infty}r^{2/\alpha}\exp(-2\alpha r)\,dr=1
\Rightarrow 
A_{\alpha}=\frac{2^{1/\alpha-1/2}\alpha^{1/\alpha+1}}
{\sqrt{\pi\Gamma(2/\alpha)}}.
\ee
 In the opposite case, when $h\gg 1$ the solution has to be close
to the oscillatory one
\be\label{oscen}
E= -\frac{1}{h}+\frac{1}{h^{3/2}}(n+1),
\ee
where $n=0,1,2...$. 
The main contribution to the normalization comes from the Gaussian part
of the wave function because the contribution of the Coulomb tails is negligible. One gets
 for the state with $n=0$ 
\be\label{Aalosc}
\varphi(r)\approx\varphi_{osc}(r)=\sqrt{\frac{\omega}{\pi}}\exp\left(-\omega r^2/2
\right),
\ee
where $\omega=1/h^{3/2}$.
Then,
\be\label{next}
A_{\alpha}=\frac{1}{\sqrt{\pi}h^{3/4}}.
\ee
The agreement between the behavior of the estimate (\ref{next}) and
the numerical calculations can be observed in the inset in Fig. 3.
  
\section{Exchange constant for $2D$ Hydrogen-like molecule}\label{ApC}

 Since $J(R)$ is exponentially small as $R\rightarrow\infty$,
$\Psi_A$ and $\Psi_S$ are solutions of the same Schr\"{o}dinger equation,
and therefore, with exponential accuracy their combinations
$$\Psi_{1,2}=\frac{\Psi_S\pm\Psi_a}{\sqrt{2}}$$ are also the solutions
of the same Schr\"{o}dinger equation with  the Hamiltonian (\ref{H1}).
They correspond to the states of "distinguishable"
particles, when, e.g. for $\Psi_1({\bf r_1, r_2})$, the first electron
is principally located near first ion at $x=-a$
and the second electron near the second ion with $x=a$.
The electron energy
$$E=-\alpha^2/2-\beta^2/2-1/R,$$
is accurate to terms $\sim 1/R^2$. Therefore we are looking for
$\Psi_{1,2}$ in a form
\bea
\Psi_1(\vec{r}_1,\vec{r}_2)&=&\phi_{\alpha}(|\vec{r}_1+\vec{a}|)
\phi_{\beta}(|\vec{r}_2-\vec{a}|)\chi(\vec{r}_1,\vec{r}_2),\nonumber\\
\Psi_2(\vec{r}_1,\vec{r}_2)&=&\phi_{\alpha}(|\vec{r}_2+\vec{a}|)
\phi_{\beta}(|\vec{r}_1-\vec{a}|)\chi(\vec{r}_2,\vec{r}_1),
\eea
where $\phi_{\alpha,\beta}$ have an asymptotic behavior 
(see Appendix \ref{ApB})
$$\varphi_{\alpha}(r)=A_{\alpha}r^{1/\alpha-1/2}e^{-\alpha r},$$
 and
$\chi$ is a slowly varying function of $\vec{r_1}, \vec{r_2}$.
 Substituting $\Psi_1$ into wave equation and neglecting the second derivatives
of $\chi$, we obtain
\be\label{chieq}
\alpha \frac{\partial\chi}{\partial x_1} -
\beta \frac{\partial \chi}{\partial x_2}
+\left [\frac{1}{\sqrt{(x_1-x_2)^2+(y_1-y_2)^2}}+\frac{1}{2a}
-\frac{1}{a-x_1}-\frac{1}{a+x_2}\right ]\chi =0.
\ee
Equation (\ref{chieq}) valid under conditions
\bea\label{chicon}
&& |x_{1,2}|\leq a, y_{12}\equiv |y_1-y_2|\ll\sqrt{a}\nonumber\\
&& R\alpha,R\beta\gg 1, R|\alpha-\beta| \ll 1.
\eea
 The general solution of (\ref{chieq}) is
\be\label{gens}
F(C_1(x_1,x_2),C_2(\chi,x_1,x_2,y_{12})=0,
\ee
where $C_1,\  C_2$ are integrals of the motion of the ordinary
differential equations:
$$ \frac{dx_1}{\alpha}=-\frac{dx_2}{\beta}=-\frac{d\chi}{\chi}
\left [\frac{1}{\sqrt{(x_1-x_2)^2+y_{12}^2}}+\frac{1}{2a}
-\frac{1}{a-x_1}-\frac{1}{a+x_2}\right ]^{-1}.
$$
 Hence
\bea\label{inmotion}
& & \frac{dx_1}{\alpha}=-\frac{dx_2}{\beta}\Rightarrow
C_1=x_1/\alpha+x_2/\beta,\\
& &\frac{dx_1}{\alpha}\left [\frac{1}{\sqrt{(x_1-x_2)^2+y_{12}^2}}+\frac{1}{2a}
-\frac{1}{a-x_1}-\frac{1}{a+x_2}\right ]=-d\ln(\chi)
\Rightarrow \nonumber\\
& & C_2=\chi\frac{\exp(x_1/2a\alpha)[a-x_1]^{1/\alpha}[a+x_2]^{1/\beta}}
{\left [\sqrt{(x_1-x_2)^2+y_{12}^2}-x_1+x_2\right ]^{\frac{1}{\alpha+\beta}}}
\eea
Combining (\ref{gens},\ref{inmotion}) we can write the function
$\chi$ as
\be\label{chisol}
\chi(x_1,x_2,y_{12})=\frac{\exp\left (-\frac{x_1}{2a\alpha}\right )
\left [\sqrt{(x_1-x_2)^2+y_{12}^2}-x_1+x_2\right ]^{\frac{1}{\alpha+\beta}}
}{[a-x_1]^{1/\alpha}[a+x_2]^{1/\beta}}f\left (\frac{x_1}{\alpha}+
\frac{x_2}{\beta}\right ),
\ee
where unknown function $f(u)$ is determined from
the fact that  $\chi\longrightarrow 1$ when $x_1\longrightarrow -a,\ x_2$
is arbitrary, or when $x_2\longrightarrow a$ and $x_1$ is arbitrary.
  Finally, after expanding $|\vec{r}\pm a|\simeq |a\pm x|+y_{12}^2/2|a\pm x|$
in the exponent, we obtain
\bea\label{wffin}
\Psi_1(\vec{r}_1,\vec{r}_2)&=&A_{\alpha}A_{\beta}\left (a+x_1\right )^
{\frac{2-\alpha}{2\alpha}}\left (a-x_2\right )^
{\frac{2-\beta}{2\beta}}\exp\left [-a(\alpha+\beta)+\beta x_2-
\alpha x_1-\frac{\alpha y_1^2}{2(a+x_1)}-\frac{\beta y_2^2}{2(a-x_2)}
\right ]\chi(\vec{r}_1,\vec{r}_2),\\
\chi(x_1,x_2,y_{12})&=&\nonumber
\eea
\be
\left\{
\begin{array}{l}
e^{-\frac{a+x_{1}}{2a\alpha}}
\left[\frac{2a}{a-x_1}\right]^{\frac{1}{\alpha}}
\alpha^{-\frac{\alpha}{(\alpha+\beta)\beta}}
\left[\frac{\beta (a+x_1)+\alpha (a+x_2)}{a+x_2}\right]^{\frac{1}{\beta}}
\left[\frac{\sqrt{(x_1-x_2)^2+y_{12}^2}-x_1+x_2}
{\sqrt{\left(\beta (a+x_1)+\alpha (a+x_2)\right)^2+(\alpha y_{12})^2}
+\beta (a+x_1)+\alpha (a+x_2)}\right]^{\frac{1}{\alpha+\beta}}
 \\
e^{-\frac{a-x_{2}}{2a\beta}}
\left[\frac{2a}{a+x_2}\right]^{\frac{1}{\beta}}
\beta^{-\frac{\beta}{(\alpha+\beta)\alpha}}
\left[\frac{\beta (a-x_1)+\alpha (a-x_2)}{a-x_1}\right]^{\frac{1}{\alpha}}
\left[\frac{\sqrt{(x_1-x_2)^2+y_{12}^2}-x_1+x_2}
{\sqrt{\left(\beta (a-x_1)+\alpha (a-x_2)\right)^2+(\beta y_{12})^2}
+\beta (a-x_1)+\alpha (a-x_2)}\right]^{\frac{1}{\alpha+\beta}}
\end{array}\right.
\ee
Here the upper expression is given for $x_1+x_2\leq 0$, and lower expression
for $x_1+x_2\geq 0$ correspondingly.

Substituting (\ref{wffin}) in Eq. (\ref{JvsInt}), and differentiating
only the exponent we obtain
\be\label{Jgen}
2J=+2(\alpha+\beta)\int_{-a}^a\left[\Psi_1\Psi_2\right]_{x_1=x_2=x}\,
d\,x\,d\,y_1d\,y_2
\ee
Introducing notations $\mu=\alpha+\beta$ and $\nu=\beta-\alpha$ and
taking into consideration that in the approximation (\ref{chicon})
$$ \sqrt{\left[(\beta+\alpha)(a-x)\right]^2+(\beta y_{12})^2}
+(\beta+\alpha)(a-x)\approx 2(\beta+\alpha)(a-x).$$
the formula (\ref{Jgen}) finally transforms to the

\be\label{Jalbet}
2J(\alpha,\beta,R)=
R^{\frac{2}{\alpha}+\frac{2}{\beta}-\frac{1}{\mu}}e^{-\mu R}
\left[D\left(\alpha,\beta,R\right)+D\left(\beta,\alpha,R\right)\right],
\ee

where $D\left(\alpha,\beta,R\right)$ is the  following function:
\bea\label{Dabr}
D\left(\alpha,\beta,R\right)&=&
4\sqrt{\pi}A_{\alpha}^2 A_{\beta}^2
\left(\frac{\mu}{2}\right)^{-1/\mu}
\Gamma\left(\frac{2+\mu}{2\mu}\right)
\left(2^{-\mu/\alpha}\frac{\mu}{\alpha}\right)^{
\frac{2\alpha}{\mu\beta}}\times\nonumber\\
& &\int_0^1\frac{
\exp\left(-(1-x)/\alpha-\nu Rx\right)
(1+x)^{2/\beta-2/\alpha+1/\mu}
(1-x)^{2/\alpha-1/\mu}
}{\left(1-x\nu/\mu\right)^{1+1/\mu}} dx.
\eea
In the case $\alpha=\beta$ it is independent of $R$ at all:
\be\label{D0a}
D_0(\alpha)\equiv 2\,D\left(\alpha,\alpha,R\right)=
8\sqrt{\pi}A_{\alpha}^4\left(\frac{1}{4\alpha}\right)^{1/2\alpha}
\Gamma\left(\frac{\alpha+1}{2\alpha}\right)
\int_0^1 \exp\left(-t/\alpha\right)
(2-t)^{1/2\alpha}t^{3/2\alpha}\,dt.
\ee

In this case

\be
\label{Jf}
2J(\alpha,\alpha,R)=D_0(\alpha)R^{7/2\alpha}\exp(-2\alpha R)
\ee

For $2D$ $H_2$ molecule ($\alpha=2,\ A_{\alpha}=4/\sqrt{2\pi}$) it gives
\be\label{H_2}
2J(2,2,R)=30.413\, R^{7/4}\exp(-4R)
\ee
For comparison we remind that in $3D$-case
$$2J_{3D}=1.636R^{5/2}\exp(-2R)$$


\begin{thebibliography}{99}
\bibitem{fit} We found that for three-dimensional hydrogen molecule
$\gamma_1=0.23$ and $\gamma_2=0.19$.
\end{thebibliography}
\end{document}